\RequirePackage{lineno}
\documentclass[prd,english,showpacs,superscriptaddress,amsmath,amssymb,floatfix,nofootinbib]{revtex4-2}

\makeatletter
\def\@fnsymbol#1{%
 \ensuremath{%
  \ifcase#1\or
   *\or                        \dagger                   \or
   \ddagger                \or \mathsection              \or
   \mathparagraph\or
   **\or                       \dagger\dagger            \or
   \ddagger\ddagger        \or \mathsection \mathsection \or
   \mathparagraph\mathparagraph\or
   *{*}*\ignorespaces      \or \dagger\dagger\dagger     \or
   \ddagger\ddagger\ddagger\or \mathsection \mathsection \mathsection \or
   \mathparagraph\mathparagraph\mathparagraph\or
  \else
   \@ctrerr
  \fi
 }%
}%
\makeatother

\usepackage{tikz}  

\usepackage{graphicx}
\usepackage[normalem]{ulem}
\usepackage[caption=false]{subfig}
\usepackage{float}
\usepackage{soul,xcolor}
\usepackage{nicefrac}
\usepackage[colorlinks = true,linkcolor = blue, urlcolor  = blue,citecolor = red,anchorcolor = blue]{hyperref}

\definecolor{darkgreen}{RGB}{50,205,50}

\let\oldcite\cite
\renewcommand{\cite}[1]{\mbox{\oldcite{#1}}}  %%% This avoids line breaks in citations
\newcommand{\xfitter}{\textsc{xFitter}}
\newcommand{\DYTurbo}{\texttt{DYTurbo}}
\newcommand{\ReneSANCe}{\textsc{ReneSANCe}}

\newcommand{\umeas}[1]{\,\text{#1}} % use this when writing unit of measurement in math mode
\newcommand{\inlinemaketitle}{{\let\newpage\relax\maketitle}}

\begin{document}

%\setpagewiselinenumbers
% \modulolinenumbers[5]
% \linenumbers
\setstcolor{blue}
\hyphenpenalty=10000
%\looseness=-2

%%%%%%%%%%%%%%%%%%%%%%%%%%%%%%%%%%%%%%
%%%%%%%%%%%%%%%%%%%%%%%%%%%%%%%%%%%%%%
%\title {Open source QCD analysis of pion fragmentation function with uncertainties in the xFitter framework}
\title {Drell-Yan cross-sections with fiducial cuts: \\ impact of linear power corrections and $q_T$-resummation in PDF determination \footnote{\it Presented at DIS2022: XXIX International Workshop on Deep-Inelastic Scattering and Related Subjects, Santiago de Compostela, Spain, May 2-6 2022.}}
 
%%%%%%%%%%%%%%%%%%%%%%%%%%%%%%%%%%%%%%%%%%%%%%%%%%%%%%%%%%%%%%%%%%%%%%%
% DEFINE AFFILIATIONS

%%%%%%%%%%%%%%%%%%%%%%%%%%%%%%%%%%%%%%%%%%%%%%%%%%%%%%%%%%%%%%%%%%%%%%%

% Hamzeh: I recommend adding the ORCID number for the authors.

\author{Simone Amoroso}
%\email{Simone.Amoroso@cern.ch}
\author{Ludovica Aperio Bella}
\author{Maarten Boonekamp}
\author{Stefano Camarda}
\author{Alexander Glazov}
\author{Alessandro Guida\footnote{Speaker \email{alessandro.guida@cern.ch}}}
\author{Renat Sadykov}
\author{Yulia Yermolchyk}

%\author{\textcolor{red}{ other xFitter names here}}

%\dubna{} \maxPlanck{} \oxford{} \rome{} \cracow{} \hamburg{}

%%%%%%%%%%%%%%%%%%%%%%%%%%%%%%%%%%%%%%
\date{\today}

\begin{abstract}
Measurement at Hadron colliders of neutral- and charged-current Drell-Yan production provide essential constraints in the determination of parton distribution functions. Experimentally, they have reached percent level precision, challenging the accuracy of the theoretical predictions.
In this work we benchmark the novel implementation in \texttt{DYTurbo} of linear fiducial power corrections in the $q_T$-subtraction formalism at NLO and NNLO in QCD. We illustrate how the inclusion of linear fiducial power corrections impacts predictions for precise $W$ and $Z$ measurements from the LHC and affects their description by modern global parton distribution functions. The further inclusion of $q_T$-resummation corrections in the theoretical predictions leads to a better modelling of the lepton $p_T$ distribution and we study how these improve the description of the data.\\
[10pt]
%
%%%%%%%%%%%%%%%%%%%%%%%%%%%%%%%%%%%%%%%%%%
% \hspace*{2cm}
% \textcolor{red}{\small  Link to the ``TO DO'' list on Google Docs is
% \href{https://docs.google.com/document/d/1sfb2d_Yvnc8LZYIz5kcuQMDY-W6Zk5R4pw0bdFfRKwA/edit?usp=sharing}{\underline{here.}}
% }%%%%%%%%%%%%%%%%%%%%%%%%%%%%%%%%55
%\vspace*{3cm}
\vspace*{1.5cm}
\vspace*{0.8cm}
\end{abstract}
%
%\pacs{13.66.Bc, 13.87.Fh, 13.85.Ni}

{
\let\clearpage\relax
\let\newpage\relax
\maketitle
}
%{\it Presented at DIS2022: XXIX International Workshop on Deep-Inelastic Scattering and Related Subjects, Santiago de Compostela, Spain, May 2-6 2022}.

%\tableofcontents{}
%%%%%%%%%%%%%%%%%%%%%%%%%%%%%%%%%%%%%%
%%%%%%%%%%%%%%%%%%%%%%%%%%%%%%%%%%%%%%
%
%\newpage
\section{Introduction} \label{s:intro}

The Drell-Yan (DY) process consists of lepton pair production, through the creation of a vector boson, either $\gamma^*/Z$ or $W^{\pm}$, in hadron-hadron high energy collisions. Measurements of this process help to probe the parton distribution functions (PDF) giving insight on the $u$- and $d$-valence quarks PDFs and the sea/light-quark decomposition.

To extract precise PDFs, the level of accuracy reached by the experiments in DY measurements needs to be matched by the precision of the theoretical predictions. 
%The prime process for precision benchmarking is the Drell-Yan (DY) process, this can be both measured and predicted with high accuracy.
%The calculations need to take into account for extra radiation in the final state with the most important contribution coming from quantum chromodynamics (QCD) radiation.
%To implement the fiducial cuts  adopted in the experiments, the fully differential cross-section for the vector boson production is needed. This is known up to NNLO in QCD for the Drell-Yan process since some years \cite{nnlo-exclusive-dy-catani,nnlo-dy-fewz,nnlo-dy-mcfm} and most recent developments are extending the calculation up to N3LO [...].
The fully differential cross section considering leptonic decay is known up to next-to-next-to-leading-order (NNLO) in Quantum Chromodynamics (QCD) \cite{nnlo-w-fully-diff,nnlo-wz-fully-diff,nnlo-exclusive-dy-catani}  and at next-to-leading-order (NLO) in Electroweak (EW) couplings \cite{nlo-ew-I,nlo-ew-II,nlo-ew-III}. More recently the N3LO QCD calculations have also been performed \cite{n3lo-I,n3lo-II,n3lo-III,n3lo-IV,https://doi.org/10.48550/arxiv.2207.07056}. 
%Furthermore, for a meaningful comparison to the experimetal data, the prediction must implement the fiducial cuts, kinematic requirements of the final state leptons, adopted in the experiments.
%The experimental results are often extracted within the fiducial region, that corresponds to the application of kinematic cuts on the final state leptons. For a meaningful comparison with the experimental data, the prediction must also implement fiducial cuts. 
%This require the knolwdge the fully exclusive cross-section for the vector boson production is needed. This is known up to NNLO in QCD for Drell-Yan process \cite{nnlo-exclusive-dy-catani,nnlo-dy-fewz,nnlo-dy-mcfm}. 
%The state of the art precision of QCD calculation ims NNLO (order $\alpha_S$) is QCD. 
%The NNLO calculation considers up to two additional emissions and must deal with the cancellation of the related soft and collinear singularities. To this purpose two main approaches have been developed. These are often referred to as \emph{slicing} and \emph{local} singularity subtraction schemes. Thanks to the success of these methods,  the NNLO QCD accuracy for DY is now the state of the art and the predictions can be obtained with various public codes \cite{nnlo-dy-fewz,nnlo-dy-mcfm,MATRIX-dy-nnlo,dyturbo}. More recently also inclusive N3LO results have been obtained \cite{dy-n3lo}.
%Despite the level of precision claimed by the NNLO codes, it has been noticed that the results differ between each other by an amount that can reach the percent level in some region of the phase space 
It has been noticed \cite{moch-benchmark-nnlo} that the results from different NNLO QCD codes  differ between each other by an amount that can reach the percent level, much higher than the expected numerical differences.
%This has been noticed for example by the ATLAS collaboration in performing the $W/Z$ cross section measurement at $7\umeas{TeV}$ \cite{atlas-w-z-7tev}. 
The disagreement is understood to be related to the presence of fiducial cuts applied to the final state leptons and the different subtraction schemes adopted in the calculations \cite{fpc-tackman-I,fpc-tackman-II}. Some cut configurations lead to a linear $q_T$ dependence of the cross section and hence induce a bias in non-local subtraction calculations \cite{dyturbo,MCFM-I,MATRIX-dy-nnlo,dyres-II}. The $q_T$ dependent term is referred to as \emph{fiducial power correction} (FPC). 
%In non-local subtraction calculations a slicing in $q_T$ is performed,in order to isolate the infrared and collinear singularities; this induces a bias in the results that is the source of the differences mentioned above. 
As a solution to this problem, it has been shown that including a $q_T$ recoil prescription, the nominal accuracy of non-local subtraction codes is recovered \cite{fpc-tackman-I,dyturbo-fpc,matrix-fpc}. 
%The recoil prescription can be obtained from the resummation program and it is implemented in public codes such as DYTurbo \cite{dyturbo-fpc} and MATRIX \cite{matrix-fpc}. 
The fixed order results, regardless of the subtraction scheme used, present anyway some instabilities due to the sensitivity to enhanced $q_T$ logarithms at small $q_T$ \cite{salam-two-body-decay}. In order to obtain a physical result, these logarithms need to be resummed to all orders \cite{qt-resummation-I,qt-resummation-parisi-pretronzio,qt-resummation-collins-soper,qt-resummation-Catani_2014}.

%The fiducial $q_T$ corrections introduce in the calculations, regardless of the subtraction scheme used, an instability due to the enhanced small $q_T$ logarithms, residuals of the divergence cancellations. Using the $q_T$ resummation results, it is possible to resum these terms to all orders in the perturbative expansion and to obtain a fully physical result.

%to be affected by the enhanced $q_T/Q$ logarithms, where $Q$ is the energy scale of the process; these terms need to be resummed in order to obtain a physical result.

In this proceeding the effects of fiducial cuts and $q_T$ resummation on DY $q_T$-inclusive cross section calculations are explored. As a benchmark scenario, we consider the ATLAS $W$ and $Z$ cross section measurement at $7\,$TeV \cite{atlas-w-z-7tev}. These experimental data are very precise and offer important constraints on the PDF determination. The predictions are evaluated with \DYTurbo{} \cite{dyturbo}, a versatile program for fast DY calculations. %calculations up to N3LL' in QCD. 
The code implements a non-local $q_T$ subtraction method and easily allows the user to include the $q_T$ recoil prescription \cite{dyturbo-fpc}, or effects from $q_T$-resummation \cite{dyturbo-n3ll}. The calculations are combined with NLO EW corrections computed with the \ReneSANCe{} code \cite{renesance}.

The document is organized as follows: in section \ref{s:simulation-setup} the simulation setup for the predictions is described, next, in section \ref{s:nnlo-results}, the results are presented with a particular focus on the effects related to the FPC. In section \ref{s:data-comparison} a quantitative comparison with the ATLAS data is carried out. The data and the predictions are then used in section \ref{s:pdf-profiling} for a PDF profiling study.
%, this quantifies the impact of this measurement on the PDF determination, with a focus, in this case, on the effects of changing the underlying theory definition.
Finally, in section \ref{s:conclusion}, conclusions and future perspective are discussed.
% The predictions are used in this same for a quantitative comparison to the ATLAS data. Furthermore, as a first step to evaluate the effects of the fiducial power correction on the PDF determination, a PDF profiling study is presented. The document is organized as follows...

%In section \ref{sec:fiducial-power-corr} the problem of linear power correction that biases the results of slicing methods, in the presence of fiducial cuts, is explored. The results of the benchmark comparison are reported in section \ref{sec:nlo-nnlo-benchmark}. The predictions are produced in the fiducial region used in the ATLAS $W/Z$ $7\umeas{TeV}$ cross section measurement \cite{atlas-w-z-7tev}. The same calculations are used in section \ref{sec:pdf-study} for a PDF study where the agreement of the predictions in combination with different PDFs and the data is assessed.

\section{Simulation setup} \label{s:simulation-setup}
Predictions are calculated with \DYTurbo{} using the $G_{\mu}$ electroweak scheme:  $G_F$, $m_W$, $m_Z$ are the input values. The Standard Model input parameters are set to the following values
\begin{equation}
  \begin{alignedat}{2}
        G_F &= 1.1663787\times10^{-5}\umeas{GeV}^{-2},\quad
         & & m_Z = 91.1876 \umeas{GeV},\\
        m_W &= 80.385 \umeas{GeV}, 
        & & \Gamma_Z = 2.4950\umeas{GeV}, \\
        \Gamma_W &= 2.091 \umeas{GeV}. & & \\
  \end{alignedat}
\end{equation}
%The CKM matrix elemets are set to
%\begin{equation}
%  \begin{alignedat}{2}
%        |V_{ud}| &= 0.97427, \quad & & |V_{cd}| = 0.2252, \\
%        |V_{us}| &= 0.2253, \quad  & & |V_{cs}| = 0.97344, \\
%        |V_{ub}| &= 0.00351, \quad & & |V_{cb}| = 0.0412 \\
%  \end{alignedat}
%\end{equation}
The input PDFs are taken from the NNPDF31$\_$nnlo$\_$as$\_$0118 set \cite{nnpdf31}. The values of the renormalization and factorization scale $\mu_R$ and $\mu_F$ are set equal to the dilepton invariant mass, $m_{\ell\ell}$.  The value of the $q_T$-slicing cut-off is set to $(q_T/m_{\ell\ell})_{\text{cut}} = 0.008$. To include the resummation effects, additional parameters are cosidered: the resummation scale, $\mu_{\text{Res}}$, is set equal to the dilepton invariant mass. Non-perturbative QCD effects at low $q_T$ are included through a Gaussian form factor in the space of the impact parameter $b$, $G^\text{{NP}}(b) = \exp(-g_1 b^2)$, with $g_1=0.8$.
NLO EW corrections are calculated with the \ReneSANCe{} program and using the same EW parameters listed above as input. These include virtual weak corrections, QED initial-state radiation and initial-final interference. The data are corrected for the final state QED radiation effects using \textsc{PHOTOS} \cite{photos} and for the $\gamma \gamma \rightarrow \ell\ell$
process contributions using the \textsc{SANC} program \cite{sanc}.

The ATLAS measurement implements symmetric cuts on the final state lepton transverse momentum, $p_{T,\,\ell/\nu}>25\umeas{GeV}$. The $W$ measurement also applies a cut on the $W$-transverse mass, $m_T>40\umeas{GeV}$, with $m^2_T = 2p_{T,\ell}\,p_{T,\nu}(1-\cos \Delta\phi_{\ell,\,\nu})$. The $Z$ cross section is measured differentially in the dilepton rapidity $|y_{\ell\ell}|$  in two channels: a central channel with leptons at central pseudorapidity, $|\eta_{\ell}|<2.5$, and a forward channel in which one of the two leptons is produced at high pseudorapidity, $2.5<|\eta_{\ell}|<4.9$. Three different mass bins around the $Z$ resonance peak are considered: $m_{\ell\ell} = [46, 66, 116, 150]\umeas{GeV}$. The $W^{\pm}$ production cross section is measured as a function of the lepton pseudorapidity $|\eta_{\ell}|$. These cut configurations induce, at least in some part of phase space, linear-$q_T$ fiducial power corrections.

The predictions are produced with high statistical accuracy; the relative statistical uncertainty is at the level of fractions of permille, completely negligible with respect to the data uncertainties and the size of the effects considered in this work.
\section{NNLO QCD results} \label{s:nnlo-results}
\begin{figure}
    \centering
    \includegraphics[width = 0.4\textwidth]{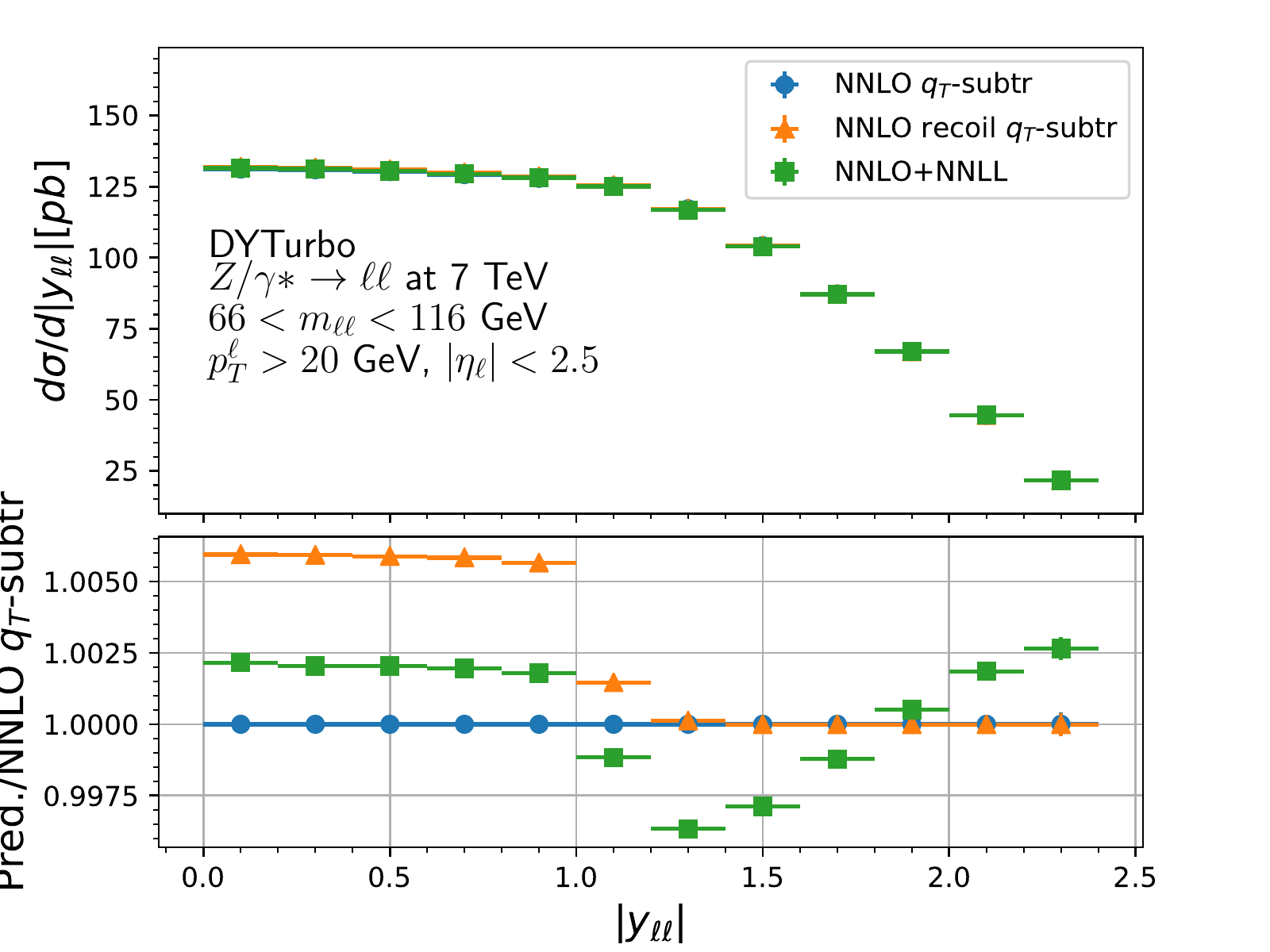}
    \includegraphics[width = 0.4\textwidth]{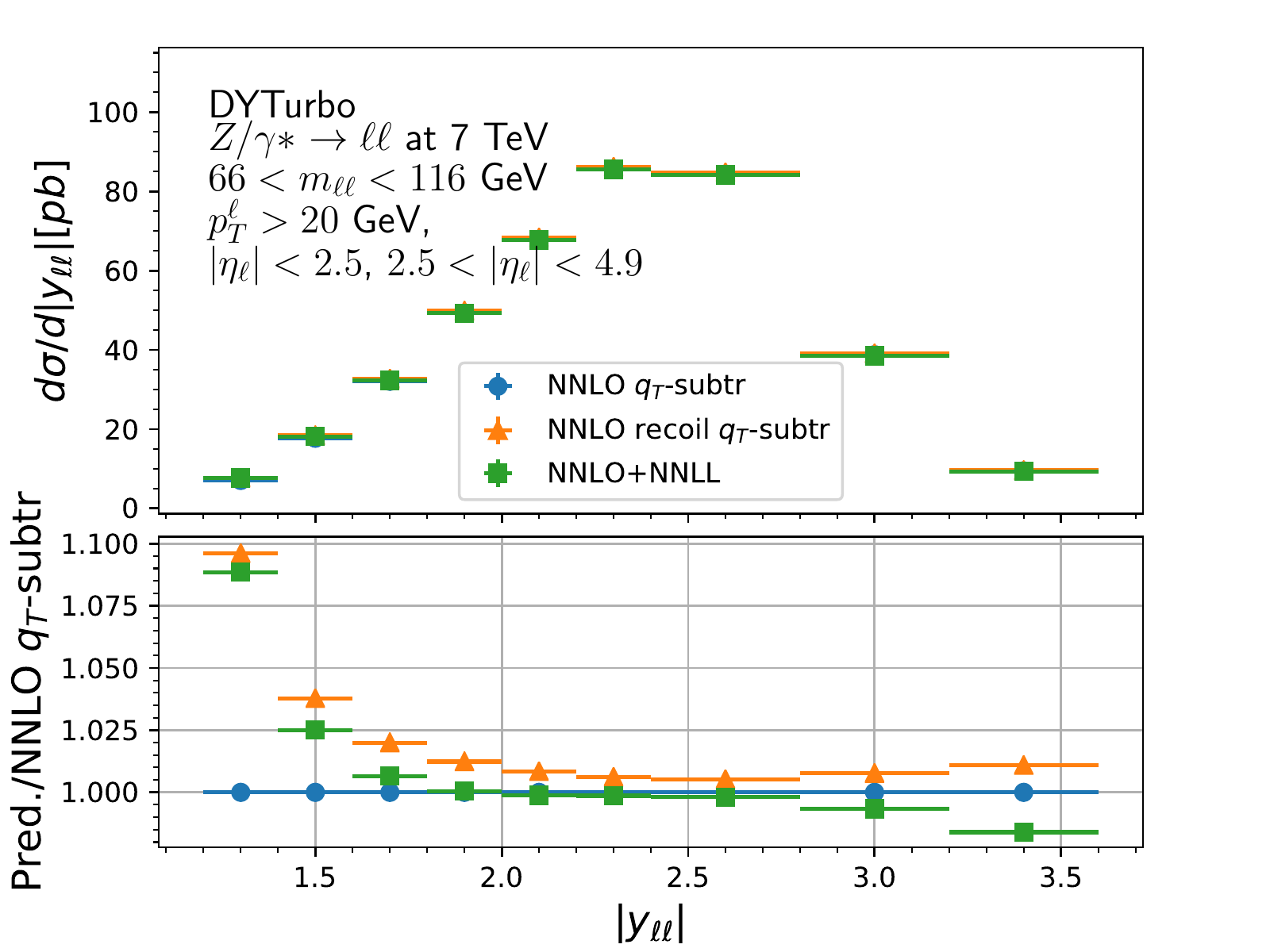}
    \caption{$Z$ boson production cross section  at NNLO (with and without implementing a $q_T$ recoil prescription) and NNLO+NNLL. Both the central channel (left) and the forward rapidity channel (right) are reported. Some significant differences between the predictions are observed in both cases.}
    \label{fig:nnlo-fo-fpc-res-zypeakcc}
\end{figure}

Three different sets of predictions at NNLO QCD are produced: the nominal fixed order one using the $q_T$ subtraction method, the fixed order including the $q_T$ recoil prescription (equivalent to a local subtraction result) and the $q_T$ resummed result (at a formal accuracy of NNLO+NNLL in QCD). The different calculations are shown in the example of the $Z$-peak bin, for the central and forward channels, in Figure \ref{fig:nnlo-fo-fpc-res-zypeakcc}. In the central channel a shape difference, between the three calculations, of $0.5\%$ is observed. In the forward channel the difference between the fixed order and the resummed prediction is more striking and gets as big as 10$\%$ in the first rapidity bin. The difference between the $q_T$ recoil calculation and the resummed one is smaller, but still of the level of few percents. 

%the bias of the fixed order $q_T$ recoil prediction can be as big as $10\%$. 
%The predictions for the $W^{\pm}$ cross section are shown if Figure \ref{fig:nnlo-fo-fpc-res-wpm}. In this case an overall $1\%$ normalization difference and a $5$ permille shape difference, in the $W^+$ case, are observed.

%\begin{figure}
    %\centering
    %\includegraphics[width = %0.49\textwidth]{NNLO_fo-fpc-res/wminus.pdf}
    %\includegraphics[width = %0.49\textwidth]{NNLO_fo-fpc-res/wplus.pdf}
    %\caption{$W^{\pm}$ boson production cross section at %NNLO (with and without implementing a $q_T$ recoil %prescription) and NNLO+NNLL. Some significant %differences between the predictions are observed}
%    \label{fig:nnlo-fo-fpc-res-wpm}
%\end{figure}
\section{Data-predictions comparison} \label{s:data-comparison}
A quantitative comparison of the predictions with the experimental data is carried out using the \xfitter{} framework \cite{xfitter-I,xfitter-II}. 
The following $\chi^2$ definition is used:
\begin{equation}\label{eq:prof-chi2}
\begin{split}
    \chi^2(\mathbf{b}_{\text{exp}}, \mathbf{b}_{\text{th}}) =&  \sum_{i=1}^{N_{\text{data}}} \frac{\left[D_i - T_i(1 - \sum_k\gamma_{ik}^{\text{th}}b_{k,\text{th}} - \sum_j\gamma_{ij}^{\text{exp}}b_{j,\text{exp}} )\right]^2}{\delta^2_{i,\text{uncor}}T^2_i+\delta^2_{i,\text{stat}}D_iT_i} \\
        & +  \sum_i \log \frac{\delta^2_{i,\text{uncor}}T_i^2 + \delta^2_{i,\text{stat}}D_iT_i}{\delta^2_{i,\text{uncor}}D_i^2 + \delta^2_{i,\text{stat}}D_i^2} \\
    & + \sum_{j=1}^{N_{\text{exp.sys}}}b^2_{j,\text{exp}} + \sum_{k=1}^{N_{\text{th.sys}}} b^2_{k,\text{th}}.
\end{split}
\end{equation}
Both the experimental uncertainties and theoretical uncertainties arising from PDF variation are considered. 
The correlated uncertainty components are accounted with two sets of nuisance parameters,
$\mathbf{b}_{\text{exp}}$ and $\mathbf{b}_{\text{th}}$. The impact of the correlated uncertainty sources on the theory point $\sigma_i^{\text{th}}$ is described by the matrices $\gamma^{\text{exp/th}}_{ij}$. $D_i$, $T_i$ are the data and theory points and  $\delta_{i,\,\text{stat}}$, $\delta_{i,\,\text{uncor}}$  are the relative statistical and uncorrelated systematic uncertainties. 

The $\chi^2$, at its minimum, provides a test of the compatibility between the data and the predictions.
The penalty term for determining the nuisance parameters is given by the last line in equation \ref{eq:prof-chi2}, this is referred to as \emph{correlated} $\chi^2$ component. The first line in the definition is instead quoted in the results as the data set $\chi^2$ component. Finally, the second line, the \emph{log penalty} term, is a small bias correction term.

Predictions for different PDFs are obtained using APPLgrids \cite{applgrid} generated at NLO QCD with MCFM \cite{MCFM-I,MCFM-II,MCFM-III}. The NNLO QCD accuracy is obtained through NNLO k-factors ($kF$) calculated using the \DYTurbo{} predictions described in section \ref{s:simulation-setup}. The $kF$ are combined multiplicatively with NLO EW $kF$ calculated with the ReneSANCE program.

As a first test, the CT14nnlo PDF set \cite{CT14-PDF} is used in the comparison. The $\chi^2$ results are reported in Table \ref{tab:chi2-data-pred-ct14nnlo}.
\begin{table}[]
    \centering
    \resizebox{0.6\textwidth}{!}{%
    \begin{tabular}{l c c c}
         & CT14nnlo $68\%$CL & & \\
         \hline
        Dataset & NNLO         & NNLO         & NNLO+ \\
                & $q_T$-subtr. & recoil       & NNLL \\
                &              & $q_T$-subtr. &  \\
        \hline
        $W^+$ lepton rapidity & $9.4/11$ & $8.8/11$ & $8.8/11$ \\
        $W^-$ lepton rapidity  & $8.2/11$ & $8.7/11$ & $8.2/11$ \\
        Low mass, Z rapidity & $11/6$ & $7.2/6$ & $7.5/6$ \\
        Mass peak, central Z rapidity & $15/12$ & $10/12$ & $7.7/12$ \\
        Mass peak, forward Z rapidity & $9.6/9$ & $5.3/9$ & $6.4/9$ \\
        High mass, central Z rapidity & $6.0/6$ & $6.5/6$ & $5.8/6$ \\
        High mass, forward Z rapidity & $5.2/6$ & $5.6/6$ & $5.3/6$ \\
        Correlated $\chi^2$ & 40 & 40 & 31 \\
        Log penalty $\chi^2$ & -4.33 & -3.39 & -4.20 \\
        \hline
        Total $\chi^2/$dof & $99/61$ & $88/61$ & $77/61$ \\
        \hline
        $\chi^2$ p-value & $0.00$ & $0.01$ & $0.08$ \\
    \end{tabular}}
    \caption{Results of the comparison of the ATLAS data \cite{atlas-w-z-7tev} with the predictions. The CT14nnlo $68\%$CL PDF is used. An improvement in the $\chi^2$ agreement is observed when including resummation effects in the predictions.}
    \label{tab:chi2-data-pred-ct14nnlo}
\end{table}
The three sets of predictions introduced in section \ref{s:nnlo-results} are used for the study. A  reduction of $\sim10$ points in the total $\chi^2$ when using a theory that include a $q_T$ recoil prescription is observed. A further improvement of additional $\sim10$ points is obtained when considering the $q_T$ resummation effects. The trend of the results is in line with the theoretical expectation of section \ref{s:intro}.
The study is extended testing other PDF sets. The total $\chi^2$ are reported in Table \ref{tab:total-chi2-pdfs}. In all the cases a similar reduction of the total $\chi^2$, of about $20-30$ when including the $q_T$-resummation, is observed.

\begin{table}[]
    \centering
    \resizebox{0.6\textwidth}{!}{%
\begin{tabular}{c | c c c}
 & Total $\chi^2$ (ndf=61) & & \\
\hline
 PDF set& NNLO & NNLO & NNLO+NLL \\
  & $q_T$ subtr. & recoil $q_T$-subtr & \\

\hline
CT10nnlo68\%CL & 100 & 85 & 76 \\
\hline
CT14nnlo68\%CL & 99 & 88 & 77 \\
\hline
CT18NNLO68\%CL & 102 & 90 & 79 \\
\hline
MMHT14nnlo68\%CL & 124 & 99 & 94 \\
\hline
NNPDF30nnlo & 139 & 133 & 111 \\
\hline
ABMP16$\_$5$\_$NNLO & 124 & 106 & 92 \\
\hline
HERAII PDF & 199 & 201 & 160 \\
\hline\hline
CT18ANNLO68 & 96 & 84 & 74 \\
\hline
MSHT20nnlo & 111 & 87 & 79 \\
\hline
NNPDF31 & 91  & 84 & 71 \\
\hline
NNPDF40nnlo & 89 & 83 & 69 \\
\end{tabular}}
    \caption{Total $\chi^2$ for the comparison between the predictions and the ATLAS data \cite{atlas-w-z-7tev}, using different PDF sets. The three different theory definitions are tested. The first half of the table considers PDFs that did not include ATLAS 7 TeV $W$ and $Z$ data, while in the second half are PDFs that used these data in their determination.}
    \label{tab:total-chi2-pdfs}
\end{table}
\section{PDF profiling}\label{s:pdf-profiling}
The values of the nuisance parameters, $\mathbf{b}_{\text{th}}$, at the $\chi^2$ minimum, equation \ref{eq:prof-chi2}, are used to obtain an optimized version of the central PDF $f'_0$
\begin{equation}
    f'_0 = f_0 + \sum_k b^{\text{min}}_{k,\text{th}}\left( \frac{f_k^+ - f_k^-}{2} + b^{\text{min}}_{k,\text{th}} \frac{f_k^+ + f_k^- -2f_0}{2} \right).
\end{equation}
Here $f_0$ is the original central PDF and $f_k^{\pm}$ are the eigenvector  up/down variation sets. Furthermore the updated PDFs have reduced uncertainties. The profiling procedure is used to test the impact of new data set on existing PDF sets. 

\begin{figure}
    \centering
    \includegraphics[width = 0.37\textwidth]{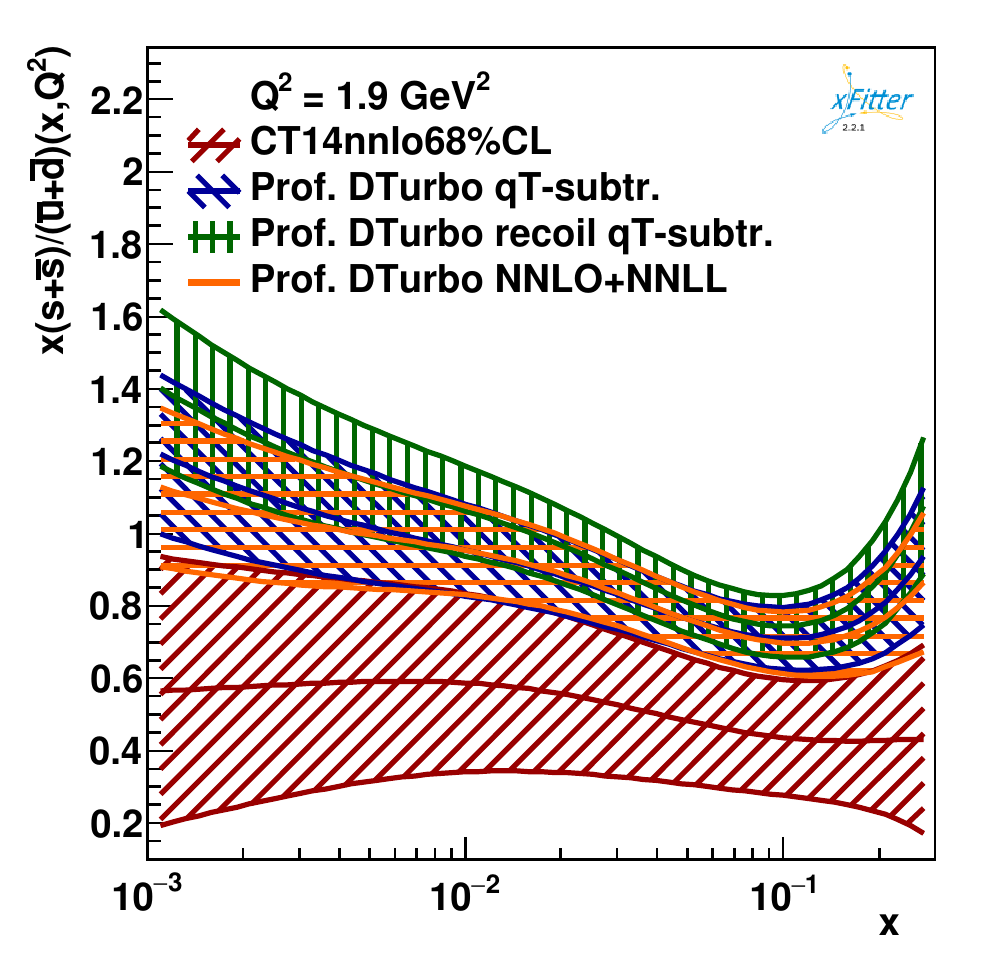}
    \includegraphics[width = 0.37\textwidth]{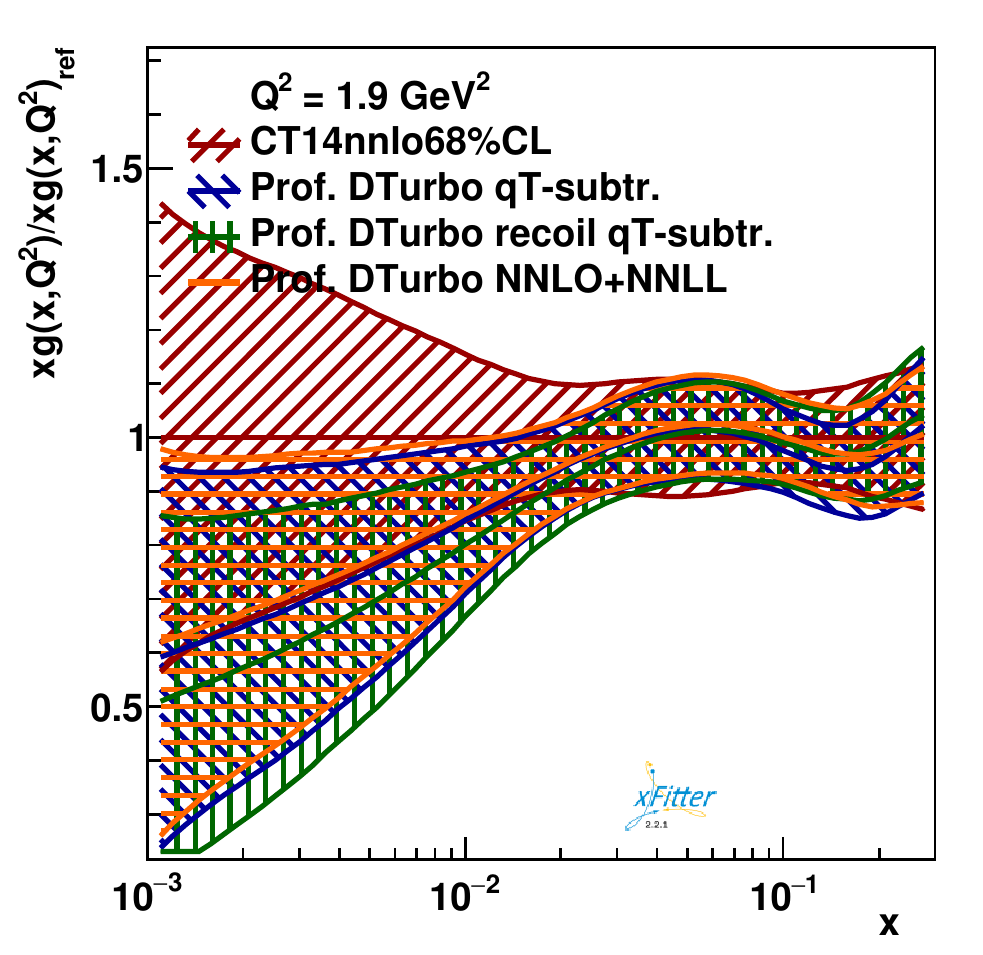}
    \caption{ Profiling results for the $R_s$  quantity (left) and the gluon PDF ratio to the CT14 set (right). The CT14 PDF set and the profiled PDF using the ATLAS data and different NNLO QCD definitions are shown. }
    \label{fig:pdf-prof-Rs-gRatio}
\end{figure}

Here the CT14nnlo 68$\%$ CL PDF set is used for the profiling; this set does not include the ATLAS $W$ and $Z$ 7 TeV data. The different NNLO(+NNLL) QCD predictions introduced in section \ref{s:nnlo-results} are used and the differences in the outcome of the profiling are examined. In Figure \ref{fig:pdf-prof-Rs-gRatio} the profiling results for two relevant quantities that are constrained by DY data are shown: the ratio $R_s = x(s+\bar{s})/(\bar{u}+\bar{d})$ and the gluon PDF. The significant impact of these data sets, as was already observed in \cite{atlas-w-z-7tev}, is clearly visible. A difference between the profiled PDFs when using the different NNLO QCD calculations is also noticeable. A general observation is that the profiled PDFs using the resummed predictions are somewhat closer to the one using the fixed order $q_T$ subtraction.

\section{Conclusion}\label{s:conclusion}
In this work the effects of fiducial cuts in the accuracy of theoretical predictions for DY cross sections has been investigated. Predictions for the ATLAS $W$ and $Z$ $7\umeas{TeV}$ cross section have been produced with the \DYTurbo{} program. The NNLO (with and without lepton $q_T$-recoil) and NNLO+NNLL in QCD accuracy, plus NLO EW corrections, have been studied. A quantitative comparison with data, including PDF uncertainties, shows a better agreement when the $q_T$ resummation effects are taken into account in the predictions. A PDF profiling procedure has been carried out to estimate the impact of using different theory definitions in PDF determinations. This exercise shows small, but noticeable differences. 
Further direction for investigation will be to study the effect of fiducial cuts on other measurement phase spaces can be investigated, and the impact on the PDF determination can be evaluated through a PDF fit to DY data.

\vspace*{2cm}
%%%%%%%%%%%%%%%%%%%%%%%%%%%%%%
%\clearpage

%\bibliographystyle{apsrev4-2}
%\bibliographystyle{apsrmp4-2}
%\bibliographystyle{aipauth4-2}
%\bibliographystyle{unsrt}
\bibliographystyle{utphys}
\bibliography{biblio}

\providecommand{\href}[2]{#2}\begingroup\raggedright\begin{thebibliography}{10}

\bibitem{nnlo-w-fully-diff}
K.~Melnikov and F.~Petriello, ``${W}$ boson production cross section at the
  large hadron collider with ${O(\alpha_S^2)}$ corrections.,''
  \href{http://dx.doi.org/10.1103/physrevlett.96.231803}{{\em Physical Review
  Letters} {\bfseries 96} no.~23, (Jun, 2006) }.
  \url{https://doi.org/10.1103%2Fphysrevlett.96.231803}.

\bibitem{nnlo-wz-fully-diff}
K.~Melnikov and F.~Petriello, ``Electroweak gauge boson production at hadron
  colliders through ${O(\alpha_S^2)}$,''
  \href{http://dx.doi.org/10.1103/physrevd.74.114017}{{\em Physical Review D}
  {\bfseries 74} no.~11, (Dec, 2006) }.
  \url{https://doi.org/10.1103%2Fphysrevd.74.114017}.

\bibitem{nnlo-exclusive-dy-catani}
S.~Catani, L.~Cieri, G.~Ferrera, D.~de~Florian, and M.~Grazzini, ``Vector boson
  production at hadron colliders: A fully exclusive {QCD} calculation at
  next-to-next-to-leading order,''
  \href{http://dx.doi.org/10.1103/physrevlett.103.082001}{{\em Physical Review
  Letters} {\bfseries 103} no.~8, (Aug, 2009) }.
  \url{http://dx.doi.org/10.1103/PhysRevLett.103.082001}.

\bibitem{nlo-ew-I}
S.~Dittmaier and M.~Krämer, ``Electroweak radiative corrections to ${W}$-boson
  production at hadron colliders,''
  \href{http://dx.doi.org/10.1103/physrevd.65.073007}{{\em Physical Review D}
  {\bfseries 65} no.~7, (Mar, 2002) }.
  \url{https://doi.org/10.1103%2Fphysrevd.65.073007}.

\bibitem{nlo-ew-II}
U.~Baur and D.~Wackeroth, ``Electroweak radiative corrections to
  ${pp/p\bar{p}\rightarrow W^{\pm}\rightarrow \ell^{\pm}\nu}$ beyond the pole
  approximation,'' \href{http://dx.doi.org/10.1103/physrevd.70.073015}{{\em
  Physical Review D} {\bfseries 70} no.~7, (Oct, 2004) }.
  \url{https://doi.org/10.1103%2Fphysrevd.70.073015}.

\bibitem{nlo-ew-III}
V.~A. Zykunov, ``{Radiative corrections to the {Drell-Yan} process at large
  dilepton invariant masses},''
  \href{http://dx.doi.org/10.1134/S1063778806090109}{{\em Phys. Atom. Nucl.}
  {\bfseries 69} (2006) 1522}.

\bibitem{n3lo-I}
C.~Duhr, F.~Dulat, and B.~Mistlberger, ``Charged current {Drell-Yan} production
  at {N3LO},'' \href{http://dx.doi.org/10.1007/jhep11(2020)143}{{\em Journal of
  High Energy Physics} {\bfseries 2020} no.~11, (Nov, 2020) }.
  \url{https://doi.org/10.1007%2Fjhep11%282020%29143}.

\bibitem{n3lo-II}
X.~Chen, T.~Gehrmann, N.~Glover, A.~Huss, T.-Z. Yang, and H.~X. Zhu, ``Dilepton
  rapidity distribution in {Drell-Yan} production to third order in {QCD},''
  \href{http://dx.doi.org/10.1103/physrevlett.128.052001}{{\em Physical Review
  Letters} {\bfseries 128} no.~5, (Feb, 2022) }.
  \url{https://doi.org/10.1103%2Fphysrevlett.128.052001}.

\bibitem{n3lo-III}
C.~Duhr and B.~Mistlberger, ``Lepton-pair production at hadron colliders at
  {N3LO} in {QCD},'' \href{http://dx.doi.org/10.1007/jhep03(2022)116}{{\em
  Journal of High Energy Physics} {\bfseries 2022} no.~3, (Mar, 2022) }.
  \url{https://doi.org/10.1007%2Fjhep03%282022%29116}.

\bibitem{n3lo-IV}
X.~Chen, T.~Gehrmann, N.~Glover, A.~Huss, P.~F. Monni, E.~Re, L.~Rottoli, and
  P.~Torrielli, ``Third-order fiducial predictions for {Drell-Yan} production
  at the {LHC},'' \href{http://dx.doi.org/10.1103/physrevlett.128.252001}{{\em
  Physical Review Letters} {\bfseries 128} no.~25, (Jun, 2022) }.
  \url{https://doi.org/10.1103%2Fphysrevlett.128.252001}.

\bibitem{https://doi.org/10.48550/arxiv.2207.07056}
T.~Neumann and J.~Campbell, ``Fiducial {Drell-Yan} production at the {LHC}
  improved by transverse-momentum resummation at {N}${^4}${LL}+{N}${^3}${LO},''
  2022.
\newblock \url{https://arxiv.org/abs/2207.07056}.

\bibitem{moch-benchmark-nnlo}
S.~Alekhin, A.~Kardos, S.~Moch, and Z.~Trócsányi, ``Precision studies for
  {Drell–Yan} processes at {NNLO},''
  \href{http://dx.doi.org/10.1140/epjc/s10052-021-09361-9}{{\em The European
  Physical Journal C} {\bfseries 81} no.~7, (Jul, 2021) }.
  \url{http://dx.doi.org/10.1140/epjc/s10052-021-09361-9}.

\bibitem{fpc-tackman-I}
M.~A. Ebert, J.~K.~L. Michel, I.~W. Stewart, and F.~J. Tackmann, ``Drell-yan
  ${q_T}$ resummation of fiducial power corrections at {N3LL},''
  \href{http://dx.doi.org/10.1007/jhep04(2021)102}{{\em Journal of High Energy
  Physics} {\bfseries 2021} no.~4, (Apr, 2021) }.
  \url{https://doi.org/10.1007%2Fjhep04%282021%29102}.

\bibitem{fpc-tackman-II}
G.~Billis, B.~Dehnadi, M.~A. Ebert, J.~K. Michel, and F.~J. Tackmann, ``Higgs
  spectrum and total cross section with fiducial cuts at third resummed and
  fixed order in {QCD},''
  \href{http://dx.doi.org/10.1103/physrevlett.127.072001}{{\em Physical Review
  Letters} {\bfseries 127} no.~7, (Aug, 2021) }.
  \url{https://doi.org/10.1103%2Fphysrevlett.127.072001}.

\bibitem{dyturbo}
S.~Camarda, M.~Boonekamp, G.~Bozzi, S.~Catani, L.~Cieri, J.~Cuth, G.~Ferrera,
  D.~de~Florian, A.~Glazov, M.~Grazzini, M.~G. Vincter, and M.~Schott,
  ``{DYTurbo}: fast predictions for {Drell–Yan} processes,''
  \href{http://dx.doi.org/10.1140/epjc/s10052-020-7757-5}{{\em The European
  Physical Journal C} {\bfseries 80} no.~3, (Mar, 2020) }.
  \url{http://dx.doi.org/10.1140/epjc/s10052-020-7757-5}.

\bibitem{MCFM-I}
J.~M. Campbell and R.~K. Ellis, ``Update on vector boson pair production at
  hadron colliders,'' \href{http://dx.doi.org/10.1103/physrevd.60.113006}{{\em
  Physical Review D} {\bfseries 60} no.~11, (Nov, 1999) }.
  \url{http://dx.doi.org/10.1103/PhysRevD.60.113006}.

\bibitem{MATRIX-dy-nnlo}
M.~Grazzini, S.~Kallweit, and M.~Wiesemann, ``Fully differential {NNLO}
  computations with {MATRIX},''
  \href{http://dx.doi.org/10.1140/epjc/s10052-018-5771-7}{{\em The European
  Physical Journal C} {\bfseries 78} no.~7, (Jun, 2018) }.
  \url{http://dx.doi.org/10.1140/epjc/s10052-018-5771-7}.

\bibitem{dyres-II}
S.~Catani, D.~de~Florian, G.~Ferrera, and M.~Grazzini, ``Vector boson
  production at hadron colliders: transverse-momentum resummation and leptonic
  decay,'' \href{http://dx.doi.org/10.1007/jhep12(2015)047}{{\em Journal of
  High Energy Physics} {\bfseries 2015} no.~12, (Dec, 2015) 1--47}.
  \url{https://doi.org/10.1007%2Fjhep12%282015%29047}.

\bibitem{dyturbo-fpc}
S.~Camarda, L.~Cieri, and G.~Ferrera, ``Fiducial perturbative power corrections
  within the ${q_T}$ subtraction formalism,''
  \href{http://dx.doi.org/10.1140/epjc/s10052-022-10510-x}{{\em The European
  Physical Journal C} {\bfseries 82} no.~6, (Jun, 2022) }.
  \url{https://doi.org/10.1140%2Fepjc%2Fs10052-022-10510-x}.

\bibitem{matrix-fpc}
L.~Buonocore, S.~Kallweit, L.~Rottoli, and M.~Wiesemann, ``Linear power
  corrections for two-body kinematics in the ${q_T}$ subtraction formalism,''
  2021.
\newblock \url{https://arxiv.org/abs/2111.13661}.

\bibitem{salam-two-body-decay}
G.~P. Salam and E.~Slade, ``Cuts for two-body decays at colliders,''
  \href{http://dx.doi.org/10.1007/jhep11(2021)220}{{\em Journal of High Energy
  Physics} {\bfseries 2021} no.~11, (Nov, 2021) }.
  \url{http://dx.doi.org/10.1007/JHEP11(2021)220}.

\bibitem{qt-resummation-I}
Y.~Dokshitzer, D.~D'yakonov, and S.~Troyan, ``On the transverse momentum
  distribution of massive lepton pairs,''
  \href{http://dx.doi.org/https://doi.org/10.1016/0370-2693(78)90240-X}{{\em
  Physics Letters B} {\bfseries 79} no.~3, (1978) 269--272}.
  \url{https://www.sciencedirect.com/science/article/pii/037026937890240X}.

\bibitem{qt-resummation-parisi-pretronzio}
G.~Parisi and R.~Petronzio, ``{Small Transverse Momentum Distributions in Hard
  Processes},'' \href{http://dx.doi.org/10.1016/0550-3213(79)90040-3}{{\em
  Nucl. Phys. B} {\bfseries 154} (1979) 427--440}.

\bibitem{qt-resummation-collins-soper}
J.~Collins, D.~E. Soper, and G.~Sterman, ``Transverse momentum distribution in
  {Drell-Yan} pair and {W} and {Z} boson production,''
  \href{http://dx.doi.org/https://doi.org/10.1016/0550-3213(85)90479-1}{{\em
  Nuclear Physics B} {\bfseries 250} no.~1, (1985) 199--224}.
  \url{https://www.sciencedirect.com/science/article/pii/0550321385904791}.

\bibitem{qt-resummation-Catani_2014}
S.~Catani, L.~Cieri, D.~de~Florian, G.~Ferrera, and M.~Grazzini, ``Universality
  of transverse-momentum resummation and hard factors at the {NNLO},''
  \href{http://dx.doi.org/10.1016/j.nuclphysb.2014.02.011}{{\em Nuclear Physics
  B} {\bfseries 881} (Apr, 2014) 414--443}.
  \url{https://doi.org/10.1016%2Fj.nuclphysb.2014.02.011}.

\bibitem{atlas-w-z-7tev}
{ATLAS collaboration}, ``Precision measurement and interpretation of inclusive
  ${W^+/W^-}$ and ${Z/\gamma^*}$ production cross sections with the {ATLAS}
  detector,'' \href{http://dx.doi.org/10.1140/epjc/s10052-017-4911-9}{{\em The
  European Physical Journal C} {\bfseries 77} no.~6, (Jun, 2017) }.
  \url{https://doi.org/10.1140%2Fepjc%2Fs10052-017-4911-9}.

\bibitem{dyturbo-n3ll}
S.~Camarda, L.~Cieri, and G.~Ferrera, ``{Drell-Yan} lepton-pair production:
  ${q_T}$ resummation at {N}${^3}${LL} accuracy and fiducial cross sections at
  {N}${^3}${LO},'' \href{http://dx.doi.org/10.1103/physrevd.104.l111503}{{\em
  Physical Review D} {\bfseries 104} no.~11, (Dec, 2021) }.
  \url{https://doi.org/10.1103%2Fphysrevd.104.l111503}.

\bibitem{renesance}
S.~Bondarenko, Y.~Dydyshka, L.~Kalinovskaya, R.~Sadykov, and V.~Yermolchyk,
  ``Hadron-hadron collision mode in {ReneSANCe}-v1.3.0,'' 2022.
\newblock \url{https://arxiv.org/abs/2207.04332}.

\bibitem{nnpdf31}
R.~D. Ball, V.~Bertone, S.~Carrazza, L.~D. Debbio, S.~Forte, P.~Groth-Merrild,
  A.~Guffanti, N.~P. Hartland, Z.~Kassabov, J.~I. Latorre, E.~R. Nocera,
  J.~Rojo, L.~Rottoli, E.~Slade, and M.~Ubiali, ``Parton distributions from
  high-precision collider data,''
  \href{http://dx.doi.org/10.1140/epjc/s10052-017-5199-5}{{\em The European
  Physical Journal C} {\bfseries 77} no.~10, (Oct, 2017) }.
  \url{http://dx.doi.org/10.1140/epjc/s10052-017-5199-5}.

\bibitem{photos}
N.~Davidson, T.~Przedzinski, and Z.~Was, ``{PHOTOS} interface in {C}++;
  technical and physics documentation,'' 2015.

\bibitem{sanc}
R.~Sadykov, A.~Arbuzov, D.~Bardin, S.~Bondarenko, P.~Christova,
  L.~Kalinovskaya, V.~Kolesnikov, A.~Sapronov, and E.~Uglov, ``{SANC} system
  and its applications for {LHC},''
  \href{http://dx.doi.org/10.1088/1742-6596/523/1/012043}{{\em Journal of
  Physics: Conference Series} {\bfseries 523} (Jun, 2014) 012043}.
  \url{https://doi.org/10.1088%2F1742-6596%2F523%2F1%2F012043}.

\bibitem{xfitter-I}
S.~Alekhin {\em et~al.}, ``{HERAFitter},''
  \href{http://dx.doi.org/10.1140/epjc/s10052-015-3480-z}{{\em Eur. Phys. J. C}
  {\bfseries 75} no.~7, (2015) 304},
  \href{http://arxiv.org/abs/1410.4412}{{\ttfamily arXiv:1410.4412 [hep-ph]}}.

\bibitem{xfitter-II}
V.~Bertone, M.~Botje, D.~Britzger, S.~Camarda, A.~Cooper-Sarkar, F.~Giuli,
  A.~Glazov, A.~Luszczak, F.~Olness, R.~Placakyte, V.~Radescu, W.~Słomiński,
  and O.~Zenaiev, ``{xFitter} 2.0.0: An open source {QCD} fit framework,''
  2017.
\newblock \url{https://arxiv.org/abs/1709.01151}.

\bibitem{applgrid}
T.~Carli, D.~Clements, A.~Cooper-Sarkar, C.~Gwenlan, G.~P. Salam, F.~Siegert,
  P.~Starovoitov, and M.~Sutton, ``A posteriori inclusion of parton density
  functions in {NLO} {QCD} final-state calculations at hadron colliders: the
  {APPLGRID} project,''
  \href{http://dx.doi.org/10.1140/epjc/s10052-010-1255-0}{{\em The European
  Physical Journal C} {\bfseries 66} no.~3–4, (Feb, 2010) 503–524}.
  \url{http://dx.doi.org/10.1140/epjc/s10052-010-1255-0}.

\bibitem{MCFM-II}
J.~M. Campbell, R.~K. Ellis, and C.~Williams, ``Vector boson pair production at
  the {LHC},'' \href{http://dx.doi.org/10.1007/jhep07(2011)018}{{\em Journal of
  High Energy Physics} {\bfseries 2011} no.~7, (Jul, 2011) }.
  \url{http://dx.doi.org/10.1007/JHEP07(2011)018}.

\bibitem{MCFM-III}
J.~M. Campbell, R.~K. Ellis, and W.~T. Giele, ``A multi-threaded version of
  {MCFM},'' 2015.

\bibitem{CT14-PDF}
S.~Dulat, T.-J. Hou, J.~Gao, M.~Guzzi, J.~Huston, P.~Nadolsky, J.~Pumplin,
  C.~Schmidt, D.~Stump, and C.-P. Yuan, ``New parton distribution functions
  from a global analysis of quantum chromodynamics,''
  \href{http://dx.doi.org/10.1103/physrevd.93.033006}{{\em Physical Review D}
  {\bfseries 93} no.~3, (Feb, 2016) }.
  \url{http://dx.doi.org/10.1103/PhysRevD.93.033006}.

\end{thebibliography}\endgroup

% %%%%%%%%%%%%%%%%%%%%%%%%%%%%%%%%%%%%%%%%%%%%%
% \clearpage{}
% % THIS GENERATES A REVTEX ERROR, BUT IT CORRECTLY PRINTS THE FIGURES AT THE END IN WIDETEXT FORMAT
% \begin{widetext}
% \printfigures
% \clearpage
% \printtables 
% \end{widetext}
%%%%%%%%%%%%%%%%%%%%%%%%%%%%%%%%%%%%5

%\printbibliography

\end{document}